\documentclass[12pt,English,
urlcolor=black,linkcolor=black]{article}

\usepackage[dvips]{graphicx}
\usepackage{epsfig}
\usepackage{amsmath,amsfonts,amssymb,amsthm}
\usepackage{mathrsfs,mathtools}
\usepackage{verbatim}
\usepackage{psfrag}
\usepackage{bm}
\usepackage{bbm}
\usepackage[utf8]{inputenc}
\usepackage[square,comma,sort&compress,numbers]{natbib}
\usepackage{color}
\usepackage{slashed}
\usepackage{upgreek}
\usepackage{enumitem}
\usepackage{float}

\usepackage[colorlinks,citecolor=black]{hyperref}

\usepackage{epsf,epsfig}
\usepackage{graphics}
\usepackage{subcaption}

\newcommand\blfootnote[1]{%
  \begingroup
  \renewcommand\thefootnote{}\footnote{#1}%
  \addtocounter{footnote}{-1}%
  \endgroup
}
\newcommand{\D}{\mathrm{d}}

\usepackage[errorstop]{feynmp}
\begin{document}
\thispagestyle{empty}

\begin{flushright}
{
\small
TUM-HEP-1180-18
}
\end{flushright}

\vspace{0.4cm}

\begin{center}
\Large\bf\boldmath
Correspondence between Thermal and Quantum Vacuum Transitions around Horizons
\unboldmath
\end{center}

\vspace{0.4cm}

\begin{center}
{Wen-Yuan Ai*\blfootnote{*~ai.wenyuan@tum.de} \\
\vskip0.4cm
{\it Physik Department T70, James-Franck-Stra{\ss}e,\\
Technische Universit\"at M\"unchen, 85748 Garching, Germany}\\
\vskip1.4cm}
\end{center}

\begin{abstract}
Recently, there are comparable revised interests in bubble nucleation seeded by black holes. However, it is debated in the literature that whether one shall interpret a static bounce solution in the Euclidean Schwarzschild spacetime (with periodic Euclidean Schwarzschild time) as describing a false vacuum decay at zero temperature or at finite temperature. In this paper, we show a correspondence that the static bounce solution describes either a thermal transition of vacuum in the static region outside of a Schwarzschild black hole or a quantum transition in a maximally extended Kruskal-Szekeres spacetime, corresponding to the viewpoint of the external static observers or the freely falling observers, respectively. The Matsubara modes in the thermal interpretation can be mapped to the circular harmonic modes from an $O(2)$ symmetry in the tunneling interpretation. The complementary tunneling interpretation must be given in the Kruskal-Szekeres spacetime because of the so-called thermofield dynamics. This correspondence is general for bubble nucleation around horizons. We propose a new paradox related to black holes as a consequence of this correspondence.
\end{abstract}

\newpage
\tableofcontents

\section{Introduction}
The discovery of Hawking radiation~\cite{Hawking:1974sw} and its consequences have clearly demonstrated that there are very deep relations between quantum theory and thermodynamics. It is now well known that there can be very different but complementary descriptions on the same physical processes near a horizon~\cite{Unruh:1976db,Israel:1976ur,Susskind:1993if}. For instance, a vacuum state viewed from an inertial observer in a flat Minkowski spacetime appears to be a thermal state to a uniformly accelerated observer (Rindler observer)~\cite{Unruh:1976db,Israel:1976ur}. A Rindler observer is associated with a local causal horizon called Rindler horizon. When applied to black hole event horizons, the complementary descriptions have in particular been incorporated into the axioms of
black hole complementarity (BHC)~\cite{Susskind:1993if}. For example, a possible virtual baryon-number-violating process of a falling proton viewed from the comoving falling observers can be a real thermally assisted process viewed from the external static observers.

While various local processes have been examined to satisfy BHC via thought experiments~\cite{Susskind:1993mu}, non-local processes such as bubble nucleation around the whole event horizon have been overlooked. Imagine, from the point of view of the freely falling observers, there is a bubble nucleation around a black hole induced by a quantum vacuum transition.\footnote{In this paper, ``quantum transition'' and ``quantum tunneling'' are used to denote exclusively the corresponding processes at zero temperature.} How would this process be described by the static observers? According to BHC, one plausible possibility is that this process corresponds to a thermal transition\footnote{We shall note that, however, by ``thermal transition'' we refer to vacuum transition at finite temperature for a quantum statistical system. That is, it is assumed that there are no contributions from classical thermodynamic fluctuations in calculating the transition rate.} viewed by the external static observers. In this paper, we shall show that this is true up to an unexpected point; in order for the external static observers and the freely falling observers to have the same transition probability, the tunneling interpretation must be given in the extended Kruskal-Szekeres spacetime. Moreover, we will show that there is always a correspondence between thermal and quantum vacuum transitions whenever the bubble is nucleated around a horizon. From this point of view, the thermal derivation in Ref.~\cite{Brown:2007sd} of the Coleman-De Luccia tunneling rate~\cite{Coleman:1980aw} is a natural consequence of this correspondence. 

On the other hand, recently there are comparable revised interests on bubble nucleation seeded by black holes for its phenomenological values in studying electroweak vacuum metastability with gravity effects~\cite{Gregory:2013hja,Burda:2015isa,Burda:2015yfa,Burda:2016mou,Tetradis:2016vqb,Gorbunov:2017fhq,Canko:2017ebb,Mukaida:2017bgd,Gregory:2018bdt}.  The measured 125 GeV Higgs boson~\cite{Aad:2012tfa,Chatrchyan:2012xdj} and the 173 GeV top-quark~\cite{Lancaster:2011wr}, suggests that the Higgs potential in the Standard Model (SM) develops a lower minimum due to the renormalization running of the Higgs self-interaction coupling. Without gravity, the lifetime of the electroweak vacuum in the SM is much longer than the age of the Universe~\cite{Isidori:2001bm,Degrassi:2012ry,Buttazzo:2013uya,Chigusa:2017dux,Andreassen:2017rzq}. However, it was shown that the false vacuum decay catalyzed by microscopic black holes could have a much bigger decay rate and for some parameter space of the coefficients of higher dimensional operators in the Higgs potential, the lifetime of the electroweak vacuum can be dramatically reduced~\cite{Burda:2015isa,Burda:2015yfa,Burda:2016mou}. Such microscopic black holes can be generated via the evaporation of primordial black holes. Since our Universe in the electroweak vacuum has enjoyed a very long safe time, the recent results imply that either the Higgs parameters are out of the relevant range or there must be very severe constraints on such microscopic black holes in the observable Universe. Further, since such relevant primordial black holes can be produced at the post-inflationary matter dominated stage, the absence of such objects would put severe constraints on inflation~\cite{Gorbunov:2017fhq}.

The process of bubble nucleation around black holes can be described by a static  bounce solution in the Euclidean Schwarzschild spacetime~\cite{Gregory:2013hja,Burda:2015yfa} (the meaning of ``static'' will be explained in Sec.~\ref{sec:Schwarzschild}). A query, however, was raised in Refs.~\cite{Gorbunov:2017fhq,Mukaida:2017bgd} that whether one shall interpret this bounce as describing a bubble nucleation in vacuum or in thermal plasma. Due to the periodicity of the Euclidean Schwarzschild time, it was conjectured that the bounce solution may actually describe bubble nucleation around a black hole surrounded by a thermal plasma. It was argued in Ref.~\cite{Gorbunov:2017fhq} that since the realistic evaporating black holes are in vacuum, the enhancement of the decay rate is doubted. In this paper, we shall show that this worry is not necessary since these two pictures, quantum vacuum transition and thermal vacuum transition, are dual with each other when the bubble is nucleated around a horizon. We will derive this correspondence in the particularly clear perturbative regime where the back-reactions to the background spacetime can be neglected. This is sufficient to clarify the following general aspect: the periodicity of the Euclidean Schwarzschild time appears because of the thermal nature of the state observed by the external static observers; and there is no necessity to fill the universe with extra plasma aside from the existing Hawking radiation in order to interpret the bounce solution. In particular, the probability calculated from the static bounce solution used in Refs.~\cite{Gregory:2013hja,Burda:2015isa,Burda:2015yfa,Burda:2016mou} gives exactly the thermal transition probability that measured by the external static observers. On the contrary, if one interprets the bounce solution as describing a quantum tunneling, one must extend the Schwarzschild spacetime to the Kruskal-Szekeres spacetime. This point leads to a theoretical paradox related to black holes as we will discuss in Sec.~\ref{sec:paradox}. It is clear that the near-horizon spacetime of a black hole is approximately a Rindler spacetime and can be most conveniently modeled in a $1+1$-dimensional setting. We therefore first demonstrate this correspondence for the false vacuum decay in the $1+1$-dimensional flat spacetime and then move to the Schwarzschild spacetime. The thermal transition is described by the imaginary-time formalism of thermal field theory which is only valid for equilibrium states. Hence, as long as the nonequilibrium effects do not cause a big deviation at the very late evaporation stage when the enhancement on the vacuum transition probability becomes significant, the calculations in Refs.~\cite{Gregory:2013hja,Burda:2015isa,Burda:2015yfa,Burda:2016mou} should be taken as reliable.     

This paper is organized as follows. In Sec.~\ref{sec:flat}, we will show the correspondence between quantum and thermal transitions of vacuum in the $1+1$-dimensional flat spacetime. In Sec.~\ref{sec:Schwarzschild}, we study this correspondence for bubble nucleation around Schwarzschild black holes. We consider the perturbative regime where the back-reactions to the background spacetime are neglected. The discussion is then extended to bubble nucleation around AdS black holes in Sec.~\ref{sec:AdS}. In Sec.~\ref{sec:paradox}, we raise a paradox from this correspondence. Discussions and conclusions are given in Sec.~\ref{sec:conc}.
Throughout of this paper, we take $\hbar=c=k_{\rm B}=1$.

\section{Vacuum Transition in $1+1$-dimensional Flat Spacetime}
\label{sec:flat}

The studies on the decay of a metastable vacuum state have been developed in the seminal works by Langer~\cite{Langer:1967ax,Langer:1969bc}, and by Coleman and Callan~\cite{Coleman:1977py,Callan:1977pt,Coleman:1988}. The descriptions on false vacuum decay at finite temperature have been given by Affleck~\cite{Affleck:1980ac} and Linde~\cite{Linde:1980tt,Linde:1981zj}. In this section, we build the correspondence between thermal and quantum vacuum transitions in the $1+1$-dimensional flat spacetime. We will first review the standard description on false vacuum decay given by Callan and Coleman~\cite{Callan:1977pt}. We emphasize that this picture of quantum tunneling at zero temperature should be taken by the inertial observers. The same result of the decay rate can be obtained from a thermal description when a Rindler frame is employed. If the spatial dimension is higher than one, the spacetime can not be foliated in a way such that we have Rindler wedges while still respect the symmetry of the bubble. Only in the $1+1$-dimensional case, this trouble disappears. But the $1+1$-dimensional Rindler spacetime shares the characters of the spacetime near an event horizon and captures the crucial elements to perform the correspondence. 

\subsection{Quantum transition for inertial observers}

\begin{figure}
\centering
\includegraphics[scale=0.6]{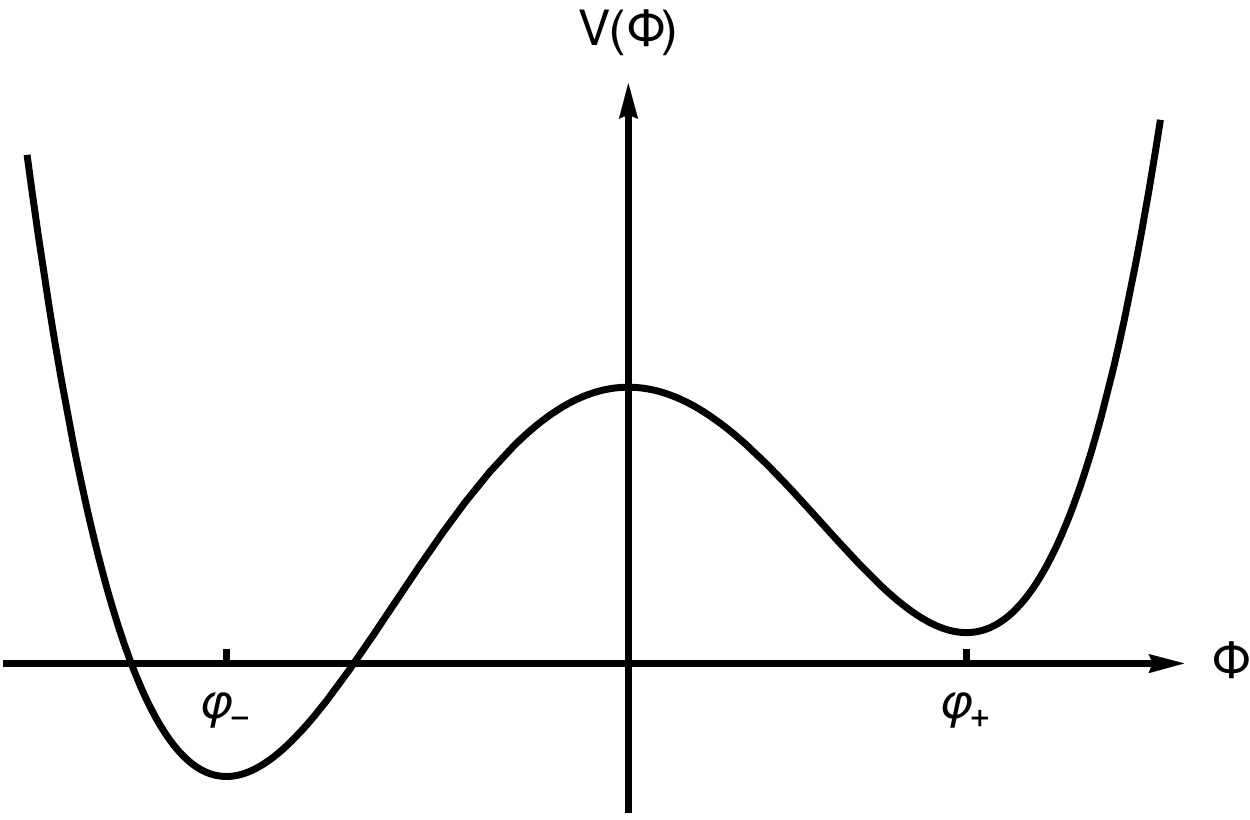}
\caption{The classical potential $V(\Phi)$ for a theory with a false vacuum.}
\label{fig:potential}
\end{figure}

Let us consider a scalar field theory  with the following action
\begin{align}
\label{Action}
S_M[\Phi]=\int\D^2x \sqrt{-\eta}\left[\frac{1}{2} \eta^{\mu\nu}(\partial_\mu\Phi)\partial_\nu\Phi-V(\Phi)\right], 
\end{align}
where $\eta$ is the determinant of $\eta_{\mu\nu}$.
In an inertial frame $\{x^0,x^1\}\equiv\{T,X\}$, the Minkowski metric $\eta_{\mu\nu}={\rm diag}(1,-1)$ .  We leave $t$ to denote the time coordinate in the Rindler frame. For clarification, we use $\Phi$ to denote the quantum field and use $\varphi$ to denote the background field.
The potential $V(\Phi)$ possesses two minima: one local $\varphi_+$ and one global $\varphi_-$, as shown in Fig.~\ref{fig:potential}. The local and global minima are called the false vacuum and the true vacuum, respectively. In the classical theory, the local minimum $\varphi_+$ is stable. However, in the quantum theory, it will decay to the global minimum through quantum tunneling, leading a bubble of true vacuum nucleated in the false vacuum. In order to describe such a tunneling process, Callan and Coleman consider the following Euclidean transition amplitude
\begin{align}
\langle\varphi_+|e^{-H\mathbb{T}}|\varphi_+\rangle,
\end{align} 
where $\mathbb{T}$ is the amount of the Euclidean time during the transition. 
Inserting a complete set of energy eigenstates and taking $\mathbb{T}\rightarrow\infty$, one obtains
\begin{align}
\label{project}
|\langle\varphi_+|0\rangle|^2\;e^{-E_0\mathbb{T}}=\int\mathcal{D}\Phi\; e^{-S_E[\Phi]}\equiv Z_E[0],
\end{align}
where 
\begin{align}
S_E[\Phi]=\int\D\mathcal{T}\D X\left[\frac{1}{2}(\partial_\mathcal{T}\Phi)^2+\frac{1}{2}(\partial_X\Phi)^2+V(\Phi)\right]
\end{align}
is the Euclidean action. We have used $\mathcal{T}$ to denote the Euclidean time in the inertial frame and leave $\tau$ to denote the Euclidean time in the Rindler frame. By calculating the path integral in Eq.~\eqref{project}, one could extract the wave function of the lowest-lying state and its energy. 

One can perform the path integral using the method of steepest descent. For the boundary conditions of interest ($\varphi(\mathcal{T}\rightarrow\pm\infty)=\varphi_+$), we can find several saddle points\footnote{\label{ft:saddle}In the strict limit $\mathbb{T}\rightarrow\infty$, there are three saddle points: the trivial false vacuum saddle point $\varphi_F(x)=\varphi_+$, the bounce $\varphi_B$ and another one $\varphi_S$ which is called the shot in Ref.~\cite{Andreassen:2016cvx}.} among which is a bounce solution $\varphi_B$. Pictorially, the bounce solution corresponds a ``particle'' initially at $\varphi_+$ rolling through the valley in $-V(\Phi)$, reaching a turning point close to $\varphi_-$ and then rolling back to $\varphi_+$---hence the name {\it bounce}. The bounce solution is responsible for the tunneling process.  It was found by Callan and Coleman that the quadratic fluctuation operator, $-\partial_\mathcal{T}^2-\partial_X^2+V''(\varphi)$, when evaluated at the saddle point $\varphi_B$, possesses a negative mode. Therefore, performing naively the Gaussian functional integral in the saddle-point expansion gives us a divergent result. This trouble reflects the fact that the state $|0\rangle$ calculated is actually metastable (cf. Eq.~\eqref{project}). The extraction of a physical result relies on a careful treatment of the analytical continuation of the path integral from a theory with only a stable vacuum to the one of interest. With this analytical continuation, one finally obtains an imaginary part in $E_0$. This imaginary part then gives the decay rate through~\cite{Callan:1977pt}\footnote{There is a sign ambiguity in extracting the imaginary part; the extraction should proceed in a way such that $\Gamma_{\rm tunn}>0$.} 
\begin{align}
\label{tun-decay}
\Gamma_{\rm tunn}=\frac{2}{\mathbb{T}}\,{\rm Im} (\ln Z_E[0]).
\end{align}
Here we would like to comment that the formula for the decay rate appears in different forms in the literature. The formula~\eqref{tun-decay} was used in Ref.~\cite{Andreassen:2016cvx}. While in Refs.~\cite{Garbrecht:2015oea,Ai:2018guc} there is not the logarithm. When there is the logarithm, one needs to take all possible multi-bounce configurations---which go back and forth from the false vacuum $N$ times, for arbitrary $N$---as (approximate) saddle points and sum over them. These multi-bounce configurations exponentiate and give the decay rate formula without the logarithm where one shall consider only the one-bounce configuration (and the trivial fasle vacuum) in evaluating the partition function~\cite{Callan:1977pt,Plascencia:2015pga}.
A study of the tunneling process in the Minkowski path integral is given in Ref.~\cite{Ai:2019fri}.

Except for the particular negative mode, the quadratic fluctuation operator evaluated at the bounce also has zero modes, originating from the spontaneous symmetry breaking of the spacetime translation symmetries caused by the bounce solution. Integrating over these zero modes also gives a divergent result. The integration over these zero modes, however, can be traded for that over the collective coordinates of the bounce, giving a factor $(V\mathbb{T})(B/2\pi)$ where $B=S_E[\varphi_B]-S_E[\varphi_+]$ and $V$ here denotes the spatial volume. Evaluating Eq.~\eqref{tun-decay} within the first quantum corrections then gives the tunneling rate per unit volume~\cite{Callan:1977pt} 
\begin{align}
\Gamma_{\rm tunn}/V=A\; e^{-B},
\end{align}  
where the prefactor $A$ takes the form ~\cite{Callan:1977pt}   
\begin{align}
\label{pref}
A=\left(\frac{B}{2\pi}\right)\left|\frac{\det'[-\Box+V''(\varphi_B)]}{\det[-\Box+V''(\varphi_+)]}\right|^{-1/2}.
\end{align}
Here $\Box=\partial_\mathcal{T}^2+\partial_X^2$ and $\det'$ implies that the zero eigenvalues of the operator $-\Box+V''(\varphi_B)$ are to be omitted when computing the determinant. Note the difference in the power of $B/2\pi$ from the $3+1$-dimensional case because we only have two zero modes in this $1+1$-dimensional case. 

\begin{figure}
\hspace{+2.5cm}
\centering
\begin{subfigure}{0.4\textwidth}
\includegraphics[scale=0.4]{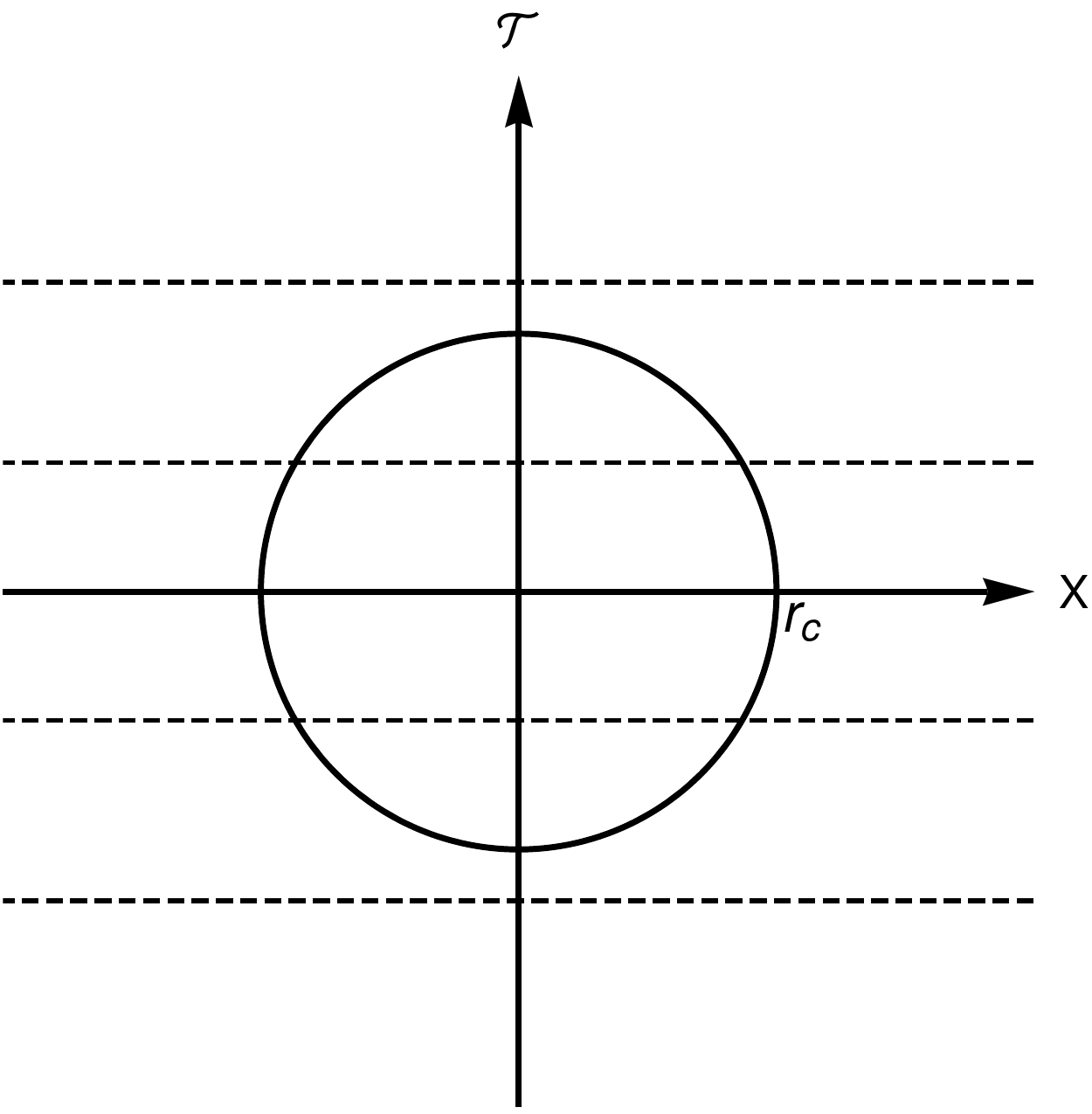}
\caption{ \hspace{+1.6cm} }
\label{fig:bounce}
\end{subfigure}
\begin{subfigure}{0.4\textwidth}
\includegraphics[scale=0.4]{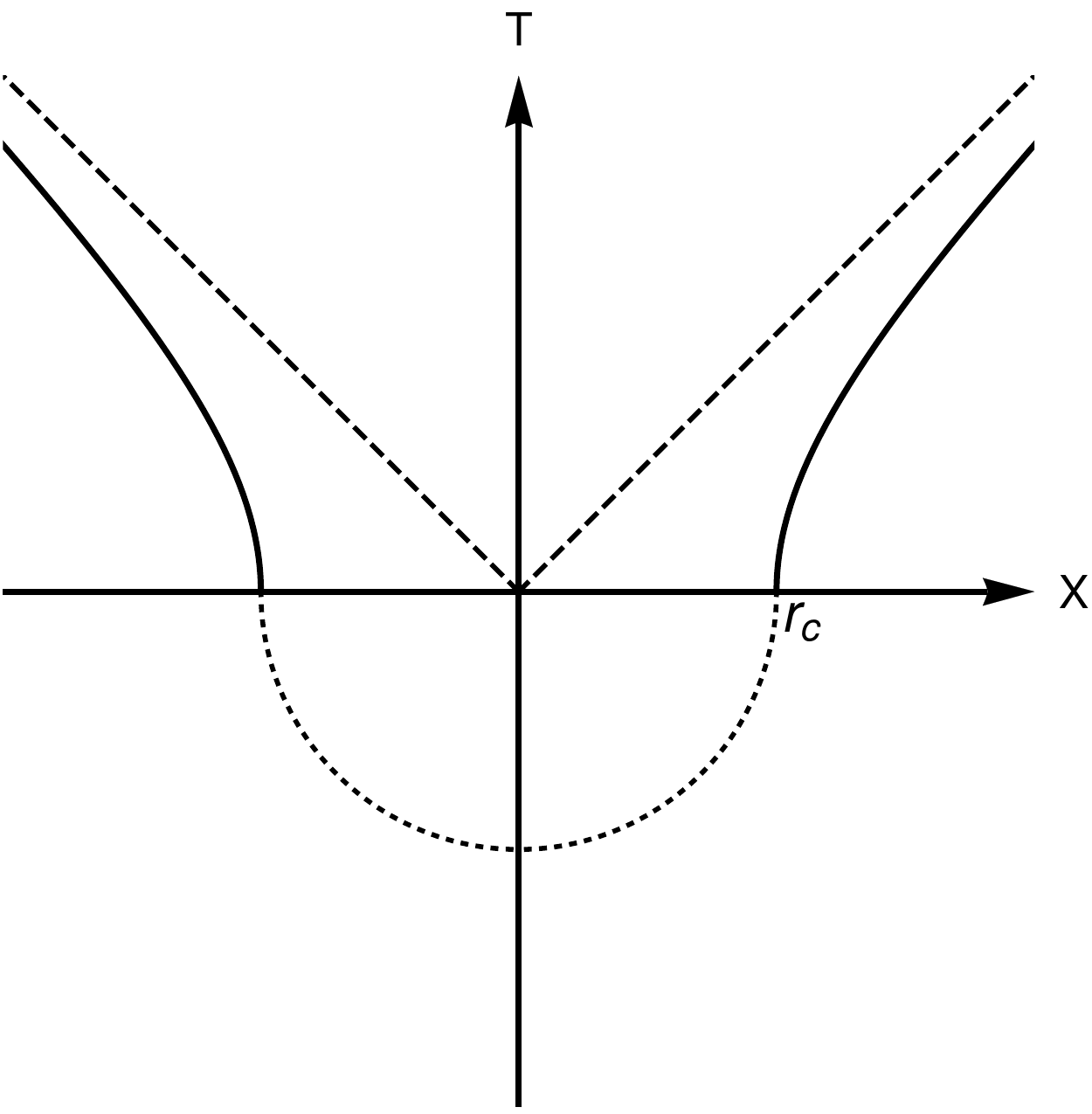}
\caption{ \hspace{+1.6cm} }
\label{fig:bubble}
\end{subfigure}
\caption{On the left panel is the bounce in the thin-wall approximation. The solid circle represents the ``bubble wall'' in the Euclidean spacetime, separating the true vacuum (inside) from the false vacuum (outside); the dashed lines represent a foliation of the Euclidean spacetime. On the right panel is the motion of the bubble wall (solid half-hyperboloid pair) in the Minkowski spacetime after the nucleation ($T\geq 0$); the dashed line is the light cone or the Rindler horizon viewed from the comoving observers with the bubble wall; the lower half-plane is continued to the Euclidean space which can be viewed as to provide the initial state at $T=0$~\cite{Hartle:1983ai,Maldacena:2001kr}.}
\label{fig:bounce&bubble}
\end{figure}

The bounce solution has an $O(2)$ symmetry and satisfies the equation of motion
\begin{align}
\label{eom1}
-\frac{\D^2\varphi}{\D r^2}-\frac{1}{r}\frac{\D\varphi}{\D r}+V'(\varphi)=0,
\end{align}
where $r=\sqrt{X^2+\mathcal{T}^2}$. The boundary conditions become $\varphi|_{r\rightarrow\infty}=\varphi_+$ and $\D\varphi/\D r|_{r=0}=0$. When the energy difference $\epsilon\equiv V(\varphi_+)-V(\varphi_-)$ is much smaller than the barrier height, we have the so-called thin-wall bounce solution. The thin-wall bounce solution takes the value $\varphi_-$ for $r$ smaller than a critical radius $r_c$ and the value $\varphi_+$ for $r$ bigger than $r_c$, with a rapid varying region around $r_c$ shown in Fig.~\ref{fig:bounce}. The bounce solution not only gives the decay rate but also gives the data of the nucleated bubble\footnote{In the $1+1$-dimensional case, the nucleated object is a kink-anti-kink pair, but we still call it bubble with a slight abuse of language.} configuration after the tunneling: $\varphi_{\rm bubble}(T=0,X)=\varphi_B(\mathcal{T}=0,X)$. The evolution of the field after the nucleation can be immediately obtained by analytically continuing the bounce solution: $\varphi_{\rm bubble}(T\geq0,X)=\varphi_{B}(\mathcal{T}\rightarrow iT,X)$. See Fig~\ref{fig:bubble} for the motion of the bubble wall. The bounce is symmetric under $\mathcal{T}\rightarrow -\mathcal{T}$. And one always cuts the bounce into two halves and analytically continues it to the Minkowski evolution at the cutting line. Such a cutting rule for the bounce is reminiscent of the optical theorem. Indeed an optical-theorem-like description of false vacuum decay is given in Ref.~\cite{Ai:2019fri}. 

\begin{figure}
\hspace{+2.0cm}
\centering
\begin{subfigure}{0.4\textwidth}
\includegraphics[scale=0.4]{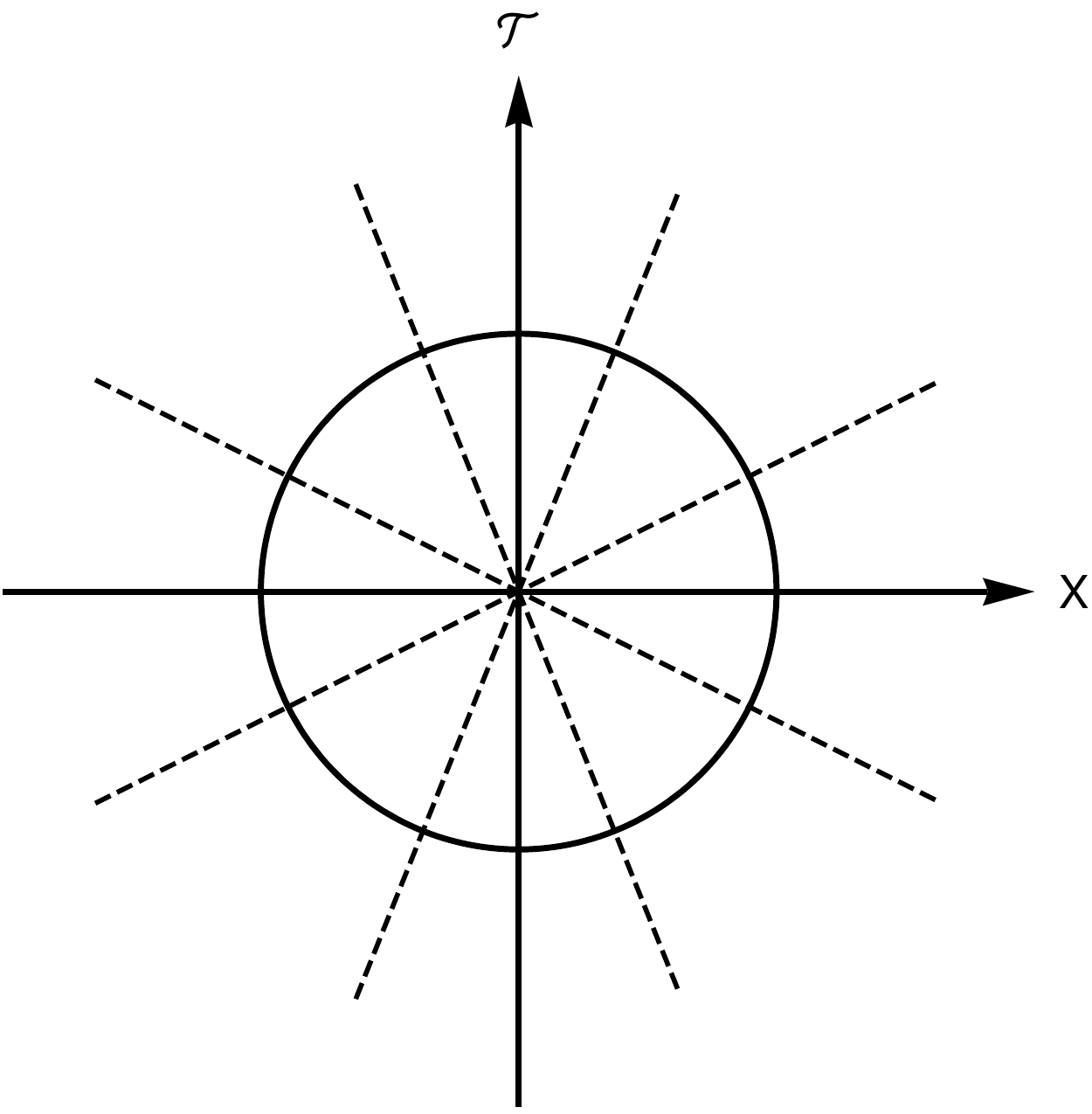}
\caption{ \hspace{+1.6cm} }
\end{subfigure}
\begin{subfigure}{0.4\textwidth}
\includegraphics[scale=0.4]{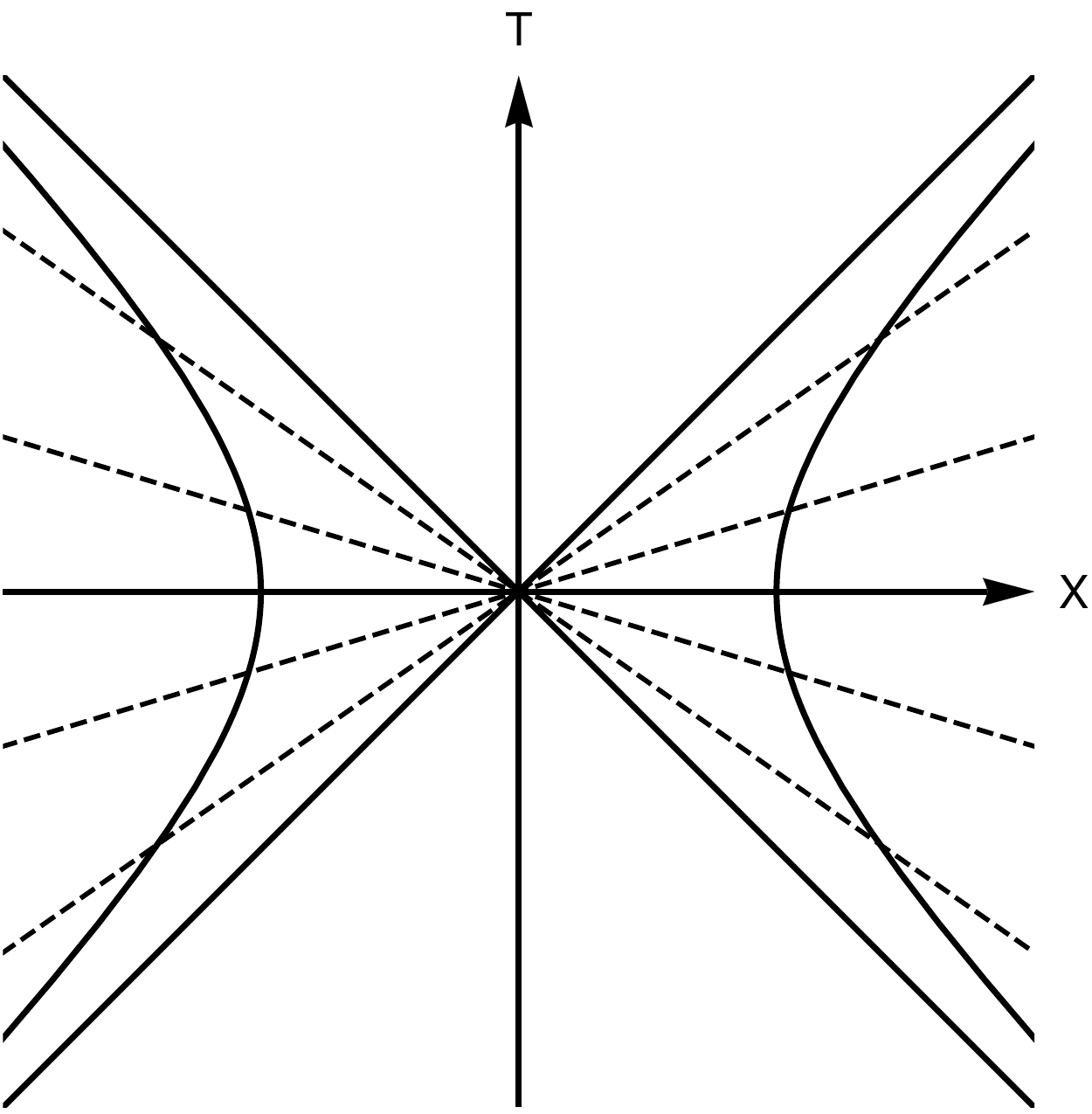}
\caption{ \hspace{+1.6cm} }
\label{fig:bubble2}
\end{subfigure}
\caption{On the left panel is a different way to do the foliation of the Euclidean spacetime; the solid circle can represent a typical worldline of an observer in this foliation; the dashed lines (the rays) represent the intermediate spatial slices and configurations.  On the right panel is the corresponding foliation in the Minkowski spacetime.}
\label{fig:bounce2}
\end{figure}

In the Callan-Coleman formalism, the bubble nucleation corresponds to a quantum tunneling process and the Euclidean time $\mathcal{T}$ is open. In this case, the bounce describes the dominant path for the transition from $\varphi_+$ at far past to $\varphi_+$ at far future in the Euclidean space. The intermediate configurations at $\mathcal{T}_i$ are given by $\varphi_B(\mathcal{T}_i,X)$ and are apparently time-dependent, as shown by the dashed lines in Fig.~\ref{fig:bounce}. It is, however, possible to do the foliation in another way as shown in Fig~\ref{fig:bounce2}. In this foliation, the spatial slices are the rays. The Euclidean time is given by the angular coordinate. Apparently, now the intermediate configurations are time-independent. This reminds us of the thermal transition at finite temperature. The observers in the Euclidean spacetime are represented by the circles which can be parameterized as $X=r\cos(\tau), \mathcal{T}=r\sin(\tau)$. If one performs an inverse Wick rotation $\tau\rightarrow it$, the circles are mapped to the hyperboloids, corresponding to uniformly accelerated observers in the Minkowski spacetime. This is in agreement with the continuation of the bubble wall trajectory.

Now one finds that the Euclidean time $\tau$ has a natural periodicity of $2\pi$ and therefore the Euclidean field theory using $\tau$ as the Euclidean time can be identified as describing a thermal state. This thermal state is the well-known Unruh effect~\cite{Unruh:1976db} felt by a uniformly accelerated observer. It should be emphasized that this thermal field theory describes only the right Rinder wedge ($X> 0, -X< T < X$; R-wedge in short) (see Fig.~\ref{fig:bubble2}) or the left Rindler wedge ($X< 0, X< T < -X$; L-wedge in short) but not both. Such kind of identification is due to thermofield dynamics~\cite{Takahasi:1974zn}.  
Thermofield dynamics describes a thermal field system by augmenting the physical Fock space by a fictitious, dual Fock space. In so doing, one can define a pure state in the doubled Fock space with the expectation value of any physical operator agreeing its statistical average in the thermal ensemble. That is to say, we can choose, for example, the R-wedge as the physical thermal system and view the L-wedge as a fictitious copy of the R-wedge. The L-wedge helps us to model the thermal state on the R-wedge from a pure state on the combination of the R- and L-wedges. The R-wedge (or the L-wedge) is quite similar to the static region outside of a black hole. We call such kind of a static region a static patch.

In conclusion, we identify two different, complementary interpretations of the Euclidean field theory in the $1+1$-dimensional flat spacetime: 
\begin{itemize}
\item[(I)] From the viewpoint of the inertial observers, it is a {\it quantum} field theory analytically continued from the original Minkowski quantum field theory formulated in the inertial frame via the Wick rotation $T\rightarrow -i\mathcal{T}$.

\item[(II)] From the viewpoint of the Rindler observers, it is a {\it thermal} field theory\footnote{Of course, by ``thermal field theory'' we actually mean thermal quantum field theory. With a slight ambiguity of the terminology, it is nevertheless clear for the differences in the two field theories interpreted in the inertial frame and the Rindler frame. The former describes a zero temperature field system while the latter describes a finite temperature one.} for the R-wedge (or another) analytically continued from the original Minkowski field theory formulated in the Rindler frame via the Wick rotation $t\rightarrow -i\tau$.
\end{itemize}
In the next subsection, we will derive the decay rate using interpretation (II).

\subsection{Thermal transition for Rindler observers}

From the relation $T=r\sinh t,\ X=r\cosh t$ (hence we choose the R-wedge as our physical thermal system), one has the metric in the coordinates $\{t,r\}$ as
\begin{align}
\D s^2=r^2\D t^2-\D r^2.
\end{align}
The action~\eqref{Action} is then expressed as
\begin{align}
S_M[\Phi]=\int\D t\int_0^\infty \D r\; r\left[\frac{1}{2r^2}(\partial_t\Phi)^2-\frac{1}{2}(\partial_r\Phi)^2-V(\Phi)\right]. 
\end{align}
The equation of motion is 
\begin{align}
\label{eom2}
\frac{1}{r^2}\frac{\partial^2\varphi}{\partial t^2}-\frac{\partial^2\varphi}{\partial r^2}-\frac{1}{r}\frac{\partial\varphi}{\partial r}+V'(\varphi)=0.
\end{align}
By a canonical transformation, we can immediately obtain the Hamiltonian in the Rindler coordinates
\begin{align}
H=\int_0^\infty\D r\; r\left[\frac{1}{2r^2}\left(\partial_t\Phi\right)^2+\frac{1}{2}\left(\partial_r\Phi\right)^2+V(\Phi)\right].
\end{align}
Taking the Wick rotation $t\rightarrow -i\tau$ and $iS_M\rightarrow -S_E$, we get 
\begin{align}
\label{therm-action}
S_E[\Phi]=\int_0^{\beta=2\pi}\D \tau\int_0^\infty \D r\; r\left[\frac{1}{2r^2}(\partial_\tau\Phi)^2+\frac{1}{2}(\partial_r\Phi)^2+V(\Phi)\right], 
\end{align}
where we have identified the periodicity of $\tau$. For convenience, we use the global Rindler temperature $T_R=1/2\pi$ rather than the proper temperature $T_P=1/(2\pi r)$ measured by the local observers.

The partition function for the thermal ensemble at temperature $1/\beta$ is given as
\begin{align}
\label{therm1+1}
Z[\beta]&={\rm Tr}[e^{-\beta H}]\notag\\
 &=\int_{\Phi(0,r)=\Phi(\beta,r)}\mathcal{D}\Phi\; e^{-\int_0^{\beta}\D \tau\int_0^\infty \D r\; r\left[\frac{1}{2r^2}(\partial_\tau\Phi)^2+\frac{1}{2}(\partial_r\Phi)^2+V(\Phi)\right]}.
\end{align} 
As explained in the last subsection, we identify Eq.~\eqref{therm1+1} as the thermal field theory for the R-wedge.
The thermal transition rate is given by 
\begin{align}
\label{therm-rate}
\Gamma_{\rm therm}=-2\,{\rm Im} F,
\end{align}
where $F$ is the free energy defined as $F=-(\ln Z[\beta])/\beta$.  

According to Linde~\cite{Linde:1980tt,Linde:1981zj}, the thermal transition can be described by a time-independent configuration in the Euclidean spacetime which is simply the solution to Eq.~\eqref{eom2} subject to the static condition $\partial\varphi/\partial t=0$ and the boundary conditions $\varphi|_{r\rightarrow\infty}=\varphi_+$, $\D\varphi/\D r|_{r=0}=0$. The false vacuum $\varphi_+$ is a trivial solution. We denote the nontrivial one as $\varphi_b$. Note that the equation of motion for this time-independent configuration $\varphi_b$ is exactly the one for the bounce (see Eq.~\eqref{eom1}). This is not surprising since the equations of motion are related by a coordinate transformation and the static condition is equivalent to the $O(2)$ symmetry of the bounce. Thus we have $\varphi_b(r)=\varphi_B(r)$. One can evaluate the partition functional $Z[\beta]$ by expanding $\Phi$ around the saddle points $\varphi_+$ and $\varphi_b$. The semiclassical result will give us an exponential of $-B$. To take the quadratic fluctuations into account, we rewrite the action~\eqref{therm-action} as
\begin{align}
\label{therm-action2}
S_E[\Phi]=\int_0^{\beta=2\pi}\D \tau\int_0^\infty \D r\; \left[-\Phi\left(\frac{1}{2r}\partial_\tau^2\right)\Phi-\Phi\left(\frac{r}{2}\partial_r^2+\frac{1}{2}\partial_r\right)\Phi+V(\Phi)\right], 
\end{align}
where the integration by parts has been used. The quadratic fluctuations at a saddle point $\varphi$ then give us the following eigenvalue problem
\begin{align}
\label{eigen}
\left[-\frac{\partial^2}{\partial r^2}-\frac{1}{r}\frac{\partial}{\partial r}-\frac{1}{r^2}\frac{\partial^2}{\partial\tau^2}+V''(\varphi)\right]\hat{\Phi}_\lambda(\tau,r)=\lambda\,\hat{\Phi}_\lambda(\tau,r).
\end{align}
One can separate the Matsubara modes as
\begin{align}
\label{decom}
\hat{\Phi}_\lambda(\tau,r)=\sum_{n=-\infty}^{n=\infty} \phi_{\lambda, n}(r) e^{i \omega_n\tau}.
\end{align}
Then we get
\begin{align}
\label{radi-eigen}
\left[-\frac{\partial^2}{\partial r^2}-\frac{1}{r}\frac{\partial}{\partial r}+\frac{\omega_n^2}{r^2}+V''(\varphi)\right]\phi_{\lambda, n}(r)=\lambda\,\phi_{\lambda, n}(r).
\end{align}
The periodic $\tau$ (recall $\tau=\tau+2\pi$) constrains $\omega_n$ to be integers. One can of course recognize Eq.~\eqref{radi-eigen} as the radial eigenvalue equation of the operator $-\Box+V''(\varphi)$ appearing in Eq.~\eqref{pref} and view Eq.~\eqref{decom} as the circular harmonic decomposition. Since we have the same saddle points (the false vacuum $\varphi_+$ and the bounce $\varphi_b=\varphi_B$), the full fluctuation spectrum in the thermal field theory $Z[\beta]$ is identical to the one  in the quantum field theory $Z_E[0]$ (cf. Eq.~\eqref{project}). It shall be straightforward to show that doing the Gaussian functional integral in $Z[\beta]$ will give us the identical result as in the quantum tunneling case.

One may still wonder how to correctly get the factor $B/2\pi$ induced by the zero modes which are not so obvious now. To see explicitly the zero modes, we define the following two independent operators
\begin{subequations}
\begin{align}
&P_X=\cos\tau\frac{\partial}{\partial r}-\frac{1}{r}\sin\tau\frac{\partial}{\partial\tau};\\
&P_{\mathcal{T}}=\sin\tau\frac{\partial}{\partial r}+\frac{1}{r}\cos\tau\frac{\partial}{\partial\tau}.
\end{align}
\end{subequations}
Acting on the equation of motion for $\varphi_b$ with these operators, one has
\begin{align}
\left[-\frac{\partial^2}{\partial r^2}-\frac{1}{r}\frac{\partial}{\partial r}+\frac{1}{r^2}+V''(\varphi_b)\right](P_{X,\mathcal{T}}\varphi_b)=0.
\end{align}
Thus we have two zero modes $B^{-1/2}P_{X,\mathcal{T}}\varphi_b$ with $\omega_n^2=1$ where we have included the normalization factor. These zero modes can be traded for the collective coordinates, giving
\begin{align}
\left(\frac{B}{2\pi}\right)\int\D X\int\D\mathcal{T}=\left(\frac{B}{2\pi}\right)\int\D\tau\int\D r\; r=\left(\frac{B}{2\pi}\right)\beta\tilde{V},
\end{align}
where we have defined the effective volume $\tilde{V}$ for the Rindler observers when we use the dimensionless time $\tau$.

Combining all the above analyses, we readily conclude that
\begin{align}
\label{corresp}
\Gamma_{\rm tunn}/V=\Gamma_{\rm therm}/\tilde{V}.
\end{align}
Dividing $\Gamma_{\rm therm}$ by $\tilde{V}$ is related to the fact that the Rindler horizon is not globally unique. One can have an infinite number of Rindler horizons with the light cone apex translated in the $X$- and $T$-directions. 
Mathematically, the result~\eqref{corresp} is quite natural since the partition function $Z[\beta]$ is related to $Z_E[0]$ simply by a coordinate transformation. Conceptually, this is nontrivial. On the LHS we describe quantum tunneling observed by the inertial observers while on the RHS we study thermal transition for Rindler observers in a Rindler wedge. For the former case, the nucleated bubble wall is represented by the paired half-hyperboloids. While for the latter case, the bubble wall is only the half-hyperboloid in the R-wedge; the other one in the L-wedge is completely fictitious. As we will se later, thermofield dynamics will have the half-hyperboloid in the L-wedge directed oppositely in time. Now we have understood that the Matsubara modes in the thermal field theory are just the circular harmonic modes in the quantum field theory. And only when we sum over all the Matsubara modes can we have agreement on the {\it exact} results on both sides.  

For general thermal vacuum transitions, the thermal transition rate at any loop order should follow from evaluating Eq.~\eqref{therm-rate}. In Refs.~\cite{Linde:1980tt,Linde:1981zj}, the one-loop vacuum decay rate at finite temperature was constructed by analogy with that at zero temperature. Such constructed one-loop formula contains functional determinants of spatial fluctuation operators, for example, the quadratic operator in Eq.~\eqref{eigen}, with the classical potential replaced by the finite temperature effective potential. Although this is definitely quantitatively correct, it has a double counting problem. The finite temperature effective potential itself comes from integrating out the thermal and quantum fluctuations while one again has the one-loop functional
determinants for the spatial fluctuation operators in the formula given in Refs.~\cite{Linde:1980tt,Linde:1981zj}. If we insist to have those functional determinants in the one-loop decay formula, the normal finite temperature effective potential should be replaced by one which is obtained by integrating out only the Matsubara modes but leaving the spatial fluctuations intact in order to avoid the double counting. We will derive such kind of effective potential in future work.

In our discussion, we evaluate Eq.~\eqref{therm-rate} by expanding around a classical static bounce $\varphi_b$. This is not possible when the finite temperature effective potential differs significantly from the classical potential such that the quantum and thermal corrected bounce can be non-perturbatively far away from the classical bounce. In this case, one must use in the expansion the quantum and thermal corrected bounce, which is obtained by solving the Schwinger-Dyson equation for the one-point function. One may then wonder whether the correspondence between thermal and quantum vacuum transitions is still valid. The answer is positive since the correspondence is really built on the map between the thermal partition function, $Z[\beta]$, and the Euclidean partition function in Eq.~\eqref{project}, $Z_E[0]$, via a coordinate transformation in the Euclidean spacetime. In particular, in the case that we worried about, we shall also evaluate $Z_E[0]$ by expanding around the quantum corrected bounce.

\section{Vacuum Transition in Schwarzschild Spacetime}
\label{sec:Schwarzschild}

Let us now study vacuum transition in the Schwarzschild spacetime. We will only consider the perturbative regime where the back-reactions to the spacetime background can be ignored.\footnote{The application condition for this perturbative regime is given in Ref.~\cite{Coleman:1980aw}.} When the background spacetime suffers a dramatic change after the bubble nucleation, the correspondence between thermal and quantum vacuum transitions has conceptual difficulties. In such a case, the remnant black hole and the original black hole can have different masses, and even worse there may not be a remnant black hole~\cite{Gregory:2013hja}. Therefore it is not clear how to define the external static observers through the bubble nucleation process in such a non-perturbative regime. Such kind of problems, however, are not exclusive for our particular correspondence between thermal and quantum vacuum transitions but are general for correspondences involved in BHC. 

We consider the following action
\begin{align}
S_M=\int\D^4x \sqrt{-g}\left[\frac{1}{2}g^{\mu\nu}(\partial_\mu\Phi)\partial_\nu\Phi-V(\Phi)\right],
\end{align}
where the metric $g_{\mu\nu}$ is given as
\begin{align}
\D s^2=\left(1-\frac{2GM}{r}\right)\D t^2-\left(1-\frac{2GM}{r}\right)^{-1}\D r^2-r^2\D\Omega_2^2,
\end{align}
with $\D\Omega_2^2\equiv\D\theta^2+\sin^2\theta\D\phi^2$ being the metric for the unit two-sphere.

We will first consider the thermal description of vacuum transitions given by the external static observers. Denoting the spatial metric as $h_{ij}=-g_{ij}$ and
\begin{align}
f(r)\equiv \left(1-\frac{2GM}{r}\right),
\end{align}
we can rewrite the action as
\begin{align}
S_M&=\int\D t\int \D^3x\sqrt{h}\left[\frac{1}{2\sqrt{f(r)}}\left(\frac{\D\Phi}{\D t}\right)^2-\frac{1}{2}\sqrt{f(r)}h^{ij}(\partial_i\Phi)\partial_j\Phi-\sqrt{f(r)}V(\Phi)\right],
\end{align}
where $h=\det(h_{ij})$.

To see that upon the Wick rotation $t\rightarrow -i\tau$ the Euclidean Schwarzschild time $\tau$ should be periodic, one can move to the Kruskal-Szekeres coordinates through ($r\geq 2GM$)
\begin{subequations}
\label{co-transf}
\begin{align}
T&=\left(\frac{r}{2GM}-1\right)^{1/2}e^{r/4GM}\sinh\left(\frac{t}{4GM}\right),\\
X&=\left(\frac{r}{2GM}-1\right)^{1/2}e^{r/4GM}\cosh\left(\frac{t}{4GM}\right)
\label{X}.
\end{align}
\end{subequations}
Note that $X\geq 0$.
The metric now reads 
\begin{align}
\label{Kruskalmetric1}
ds^2=\frac{32 G^3M^3}{r}e^{-r/2GM}(dT^2-dX^2)-r^2d\Omega_2^2,
\end{align} 
where 
\begin{align}
\label{relation}
T^2-X^2=\left(1-\frac{r}{2GM}\right)e^{r/2GM}.
\end{align}
The coordinate transformation~\eqref{co-transf} is analogous to the one from $\{t,r\}$ to $\{T,X\}$ in the $1+1$-dimensional flat spacetime. One can see that, after the Wick rotation $t\rightarrow -i\tau$, the Euclidean time $\tau$ has a natural periodicity $\beta=8\pi GM$.

Taking the Wick rotation $t\rightarrow -i\tau;\ iS_M\rightarrow -S_E$ and the identification $\tau=\tau+8\pi GM$, we have
\begin{align}
\label{action-schwarz-therm}
S_E&=\int_0^{\beta=8\pi GM}\D\tau\int \D^3x\sqrt{h}\left[\frac{1}{2\sqrt{f(r)}}\left(\frac{\D\Phi}{\D \tau}\right)^2+\frac{1}{2}\sqrt{f(r)}h^{ij}(\partial_i\Phi)\partial_j\Phi+\sqrt{f(r)}V(\Phi)\right].
\end{align}
We can identify the part of the spatial integral as the Hamiltonian $H$ with $\tau$ replaced by $t$. The thermal field theory is given by the following partition function 
\begin{align}
\label{par}
Z[\beta]={\rm Tr}[e^{-\beta H}].
\end{align}
Because of the $O(3)$ symmetry of the spacetime, we expect that the time-independent configuration responsible for the thermal transition has no dependence on $\theta$ and $\phi$. Then the action~\eqref{action-schwarz-therm} gives the equation of motion
\begin{align}
\label{eom-sch-ther}
-\frac{\D^2\varphi}{\D r^2}-\frac{f'(r)}{f(r)}\frac{\D\varphi}{\D r}-\frac{2}{r}\frac{\D\varphi}{\D r}+\frac{V'(\varphi)}{f(r)}=0.
\end{align}

To prepare for the construction of the correspondence, we can define a Euclidean ``inertial coordinate system'' $\{\mathcal{T},X\}$ through the transformation
\begin{subequations}
\label{co-transf2}
\begin{align}
\mathcal{T}&=\left(\frac{r}{2GM}-1\right)^{1/2}e^{r/4GM}\sin\left(\frac{\tau}{4GM}\right),\\
X&=\left(\frac{r}{2GM}-1\right)^{1/2}e^{r/4GM}\cos\left(\frac{\tau}{4GM}\right).
\end{align}
\end{subequations}
Note that, once we allow $\tau$ to take the whole region from $0$ to $8\pi GM$, we have extended the $X$ from $X\geq 0$ to the whole real line. But we still view Eq.~\eqref{par} as describing the thermal ensemble on the original static patch outside of the black hole.\footnote{An insignificant point: if the thermal field theory describes the static patch outside of the black hole, i.e., $r>2GM$, we shall subtract the point $X=\mathcal{T}=0$ in the Euclidean spacetime. But we can include the horizon $r=2GM$ into the system to fill this hole.} This is of course again due to thermofield dynamics which doubles the Fock space of the thermal field system. The Euclidean metric in the coordinates $\{\mathcal{T},X\}$ is 
\begin{align}
\label{Kruskalmetric2}
ds^2=\frac{32 G^3M^3}{r}e^{-r/2GM}(d\mathcal{T}^2+dX^2)+r^2d\Omega_2^2.
\end{align}

\begin{figure}
\centering
\includegraphics[scale=0.4]{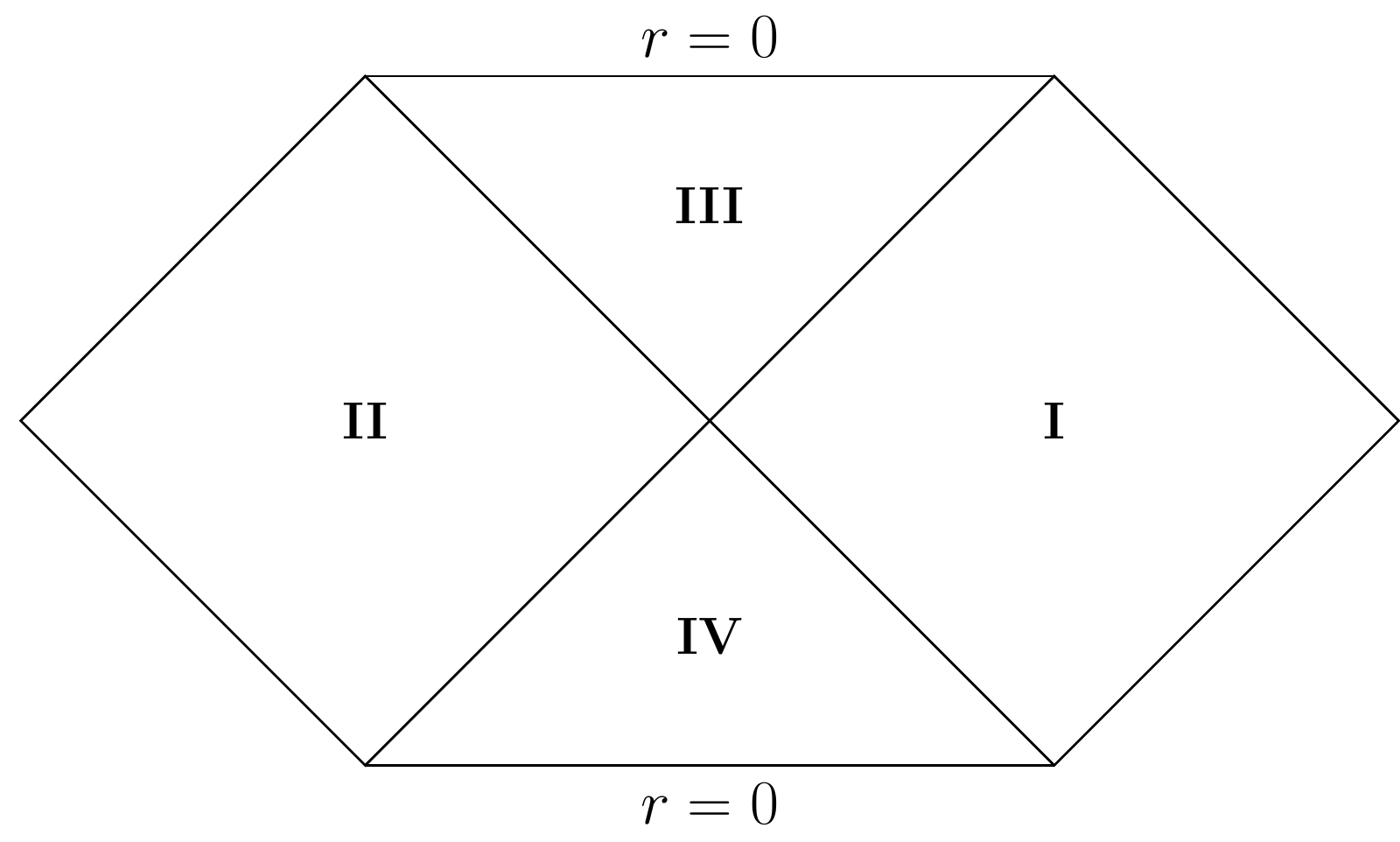}
\caption{Penrose diagram of the Kruskal-Szekeres spacetime which is obtained by a conformal transformation of the metric~\eqref{Kruskalmetric1}. The left and right square regions are two causally uncorrelated universes. The upper and lower triangle regions are the black hole and white hole, respectively. The angular directions ($\theta$ and $\phi$) are suppressed.}
\label{fig:KS}
\end{figure}

On the other hand, we can from the beginning work in the maximally extended Kruskal-Szekeres spacetime with the same metric~\eqref{Kruskalmetric1} and relation~\eqref{relation}. In this extended spacetime, there is no particular constraint on $X$ except for that $r>0$. If one ignores the unphysical white hole region, one can view this spacetime as describing two black holes in two causally uncorrelated spatial regions connected by a wormhole (the Einstein-Rosen bridge~\cite{Einstein:1935tc}); the wormhole is the shared interior of these two black holes, see Fig.~\ref{fig:KS}. Now one can study the quantum vacuum transition around the wormhole in such a spacetime. 

We shall emphasize that there is nothing singular near the wormhole event horizon and in particular, the spacetime curvature can be very small for large black holes. Therefore, the Callan and Coleman's description on false vacuum decay should be still reliable. Now we need to perform the Wick rotation $T\rightarrow -i\mathcal{T}$ which gives us the metric~\eqref{Kruskalmetric2}. The surprising thing is that after this rotation, the relation~\eqref{relation} becomes
\begin{align}
\label{relation2}
\mathcal{T}^2+X^2=\left(\frac{r}{2GM}-1\right)e^{r/2GM}\equiv\rho^2\geq 0.
\end{align}
That is, the Euclidean Kruskal-Szekeres spacetime has $r\geq 2GM$ and coincides completely with the one where the thermal field theory lives.\footnote{Because of this, there is no difference between the {\it Euclidean} Kruskal-Szekeres spacetime and the {\it Euclidean} Schwarzschild spacetime (with the Euclidean Schwarzschild time $\tau$ taking a full periodicity). Thus these two terms are used interchangeably in this paper. Without the ajective ``Euclidean'', these two spacetimes are very different.} This point also makes the Euclidean Kruskal-Szekeres spacetime be non-singular in the whole region $-\infty<\mathcal{T}<\infty; -\infty<X<\infty$ and motivates Hawking to propose the Euclidean Kruskal-Szekeres spacetime as the simplest gravitational instanton~\cite{Hawking:1976jb}.  

Using the coordinates $\{\mathcal{T},X,\theta,\phi\}$, one can have a Euclidean field theory with an open Euclidean time and the partition function\footnote{Note that the worldlines with constant $\{X,\theta,\phi\}$ are not geodesics and hence cannot represent the freely falling observers directly. But the Kruskal-Szekeres coordinates provide a very useful global coordinate frame that nevertheless can describe the quantum transition because: (i) the Euclidean Kruskal-Szekeres time is open; (ii) the bounce action, or more generally, the partition function~\eqref{qu-par} is invariant under coordinate transformations.}
\begin{align}
\label{qu-par}
Z_E[0]=\int\mathcal{D}\Phi\;e^{-S_{E,\mathcal{T}}},
\end{align} 
where we put a subscript $\mathcal{T}$ on $S_E$ to remind us of the Euclidean time we are using.
The bubble nucleation around the wormhole can be described by a bounce solution centered at the origin ($\mathcal{T}=X=0$) in the Euclidean Kruskal-Szekeres spacetime. We would like to emphasize that the coordinates $\{\mathcal{T},X\}$ here are quite similar to the one in the $1+1$-dimensional Euclidean flat spacetime. In particular, the manifolds described by $\{\mathcal{T},X\}$ are both $\mathbb{R}^2$ and the metrics possess an $O(2)$ symmetry. This is why the thermal properties of the event horizon can be understood from the Rindler horizon in $1+1$ flat spacetime. Considering the additional $O(3)$ symmetry, we expect that the bounce has an $O(3)\times O(2)$ symmetry and thus only depends on the radial distance $\rho$ (cf. Eq.~\eqref{relation2}). Such kind of bounce is called {\it static} and gives the minimal bounce action $B$~\cite{Gregory:2013hja}. Because there is no dependence on $\theta$ and $\phi$ in the static bounce solution, our tunneling problem is analogous to the one in the $1+1$-dimensional flat spacetime. For instance, it is completely trivial to check that the equation of motion for this $O(3)\times O(2)$ bounce is exactly the same as Eq.~\eqref{eom-sch-ther}. And we immediately get the agreement on the semiclassical suppressions in the thermal interpretation and the tunneling interpretation. The ``thin-wall'' bounce is described in Fig.~\ref{fig:bounce} with the extra $O(3)$ symmetry given by the $\theta$- and $\phi$-independences. There is one difference compared with the case of $1+1$-dimensional flat spacetime. In the Euclidean Kruskal-Szekeres spacetime, we do not have $\mathcal{T}$- and $X$-translation symmetries any more. Hence we shall consider the total quantum transition probability instead of the transition rate. We have
\begin{align}
P_{\rm tunn}=2\,{\rm Im}(\ln Z_E[0]).
\end{align} 
Similarly we have the thermal transition probability
\begin{align}
P_{\rm therm}=2\, {\rm Im}(\ln Z[\beta]).
\end{align}

As before, one can further perform the Gaussian functional integrals in the saddle-point expansions. Since the action $S_{E,\mathcal{T}}$ in the quantum field theory (cf. Eq.~\eqref{qu-par}) is related to the action in the thermal field theory (Eq.~\eqref{action-schwarz-therm}) by a coordinate transformation, the fluctuation spectra in both theories must match. The Matsubara modes in the thermal field theory are mapped to the circular harmonic modes (from the $O(2)$ in the $O(3)\times O(2)$) in the quantum field theory. Therefore the thermal transition probability and the quantum transition probability will have identical results: $P_{\rm tunn}=P_{\rm therm}$.  

\begin{figure}
\centering
\includegraphics[scale=0.3]{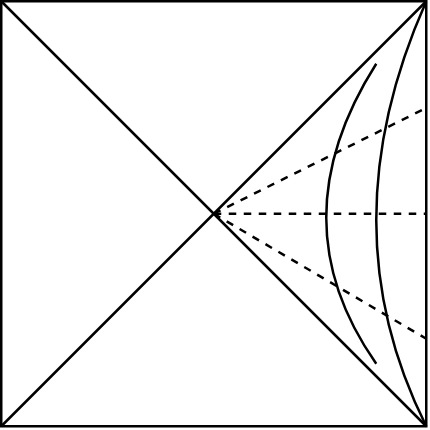}
\caption{Penrose diagram of the de Sitter space. The right triangle region is one of the static patches. The solid curved lines represent static observers. The dashed lines are spatial slices of constant static time in the static coordinates.}
\label{fig:dS}
\end{figure}

One can easily extend this correspondence between thermal and quantum vacuum transitions to other horizons, e.g., to the de Sitter horizon. The de Sitter spacetime has two causally uncorrelated static patches which are separated by de Sitter horizons, see Fig.~\ref{fig:dS}. See also Ref.~\cite{Spradlin:2001pw} for more details. Since the de Sitter spacetime is maximally symmetric, we can always choose one Kruskal (or static) coordinate system such that the bubble is nucleated around a de Sitter horizon. Viewed from the static observers in this particular coordinate system, the bubble nucleation is thermal. By using a WKB approach, Brown and Weinberg~\cite{Brown:2007sd} showed the agreement on the semiclassical suppression in the transition rates from a thermal transition description and the Coleman-De Luccia tunneling prescription~\cite{Coleman:1980aw}. However, by repeating the analysis we have given, one can show the exact agreement in the transition rates beyond the semiclassical level. 
 
\section{Bubble Nucleation around AdS Black Holes}
\label{sec:AdS}

It is well known that a physical system with gravity may be holographically dualized to a field system on its boundary~\cite{tHooft:1993dmi,Susskind:1994vu}. This holographic principle is mostly rigorously understood in the AdS/CFT duality~\cite{Maldacena:1997re,Witten:1998qj,Gubser:1998bc}. Although the constructed correspondence between thermal and quantum vacuum transitions, as other possible correspondences described by 
BHC, is quite general, it is beneficial to look into an example where the AdS/CFT duality applies. In this section, we briefly comment on bubble nucleation around eternal AdS-Schwarzschild black holes.

\begin{figure}
\centering
\includegraphics[scale=1]{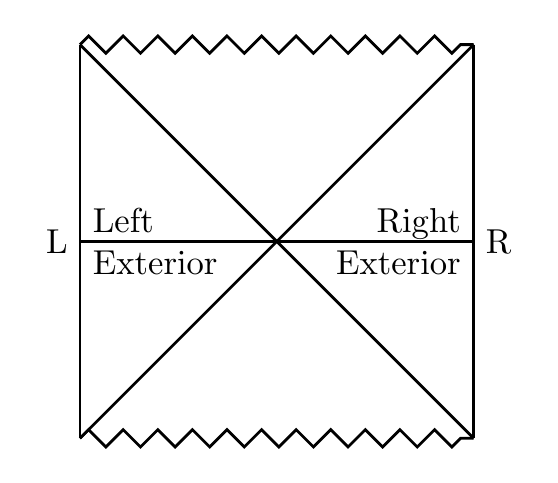}
\caption{Penrose diagram of the eternal AdS-Schwarzschild black hole. Note that the two boundaries, R and L, have topology $\mathbb{R}\times S^{D-1}$ where $D$ is the dimension of the bulk spacetime.}
\label{fig:eternal-Penrose}
\end{figure}

An eternal black hole has an extended Penrose diagram as shown in Fig.~\ref{fig:eternal-Penrose}. This Penrose diagram has two asymptotic AdS exterior regions, denoted as ``Left Exterior'' and ``Right Exterior'', which are causally uncorrelated. Correspondingly, the conformal field theory (CFT) that describes this spacetime has two identical uncoupled sectors, L and R. The eternal black hole is described by the entangled state~\cite{Maldacena:2001kr}
\begin{align}
|\Psi\rangle=\sum_n e^{-\beta E_n/2}|n\rangle_L\otimes|n\rangle_R,
\end{align}
where $|n\rangle_{L,R}$ are the energy eigenstates of energy $E_n$ of the CFT in the left and right sectors, respectively. $\beta$ is the inverse temperature of the black hole.

\begin{figure}
\centering
\hspace{-0.5cm}
\begin{subfigure}{0.25\textwidth}
\includegraphics[scale=0.8]{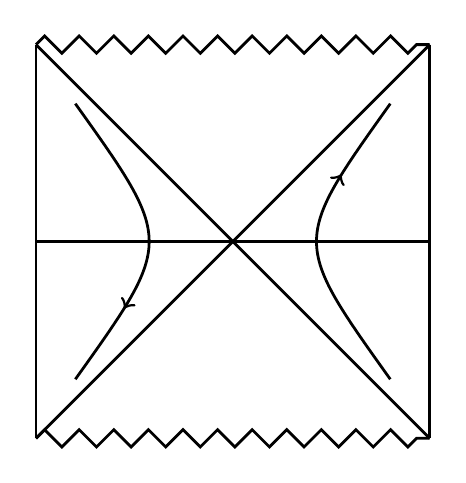}
\caption{}
\label{fig:eternal-one-BH}
\end{subfigure}
\hspace{+1.8cm}
\begin{subfigure}{0.25\textwidth}
\includegraphics[scale=0.8]{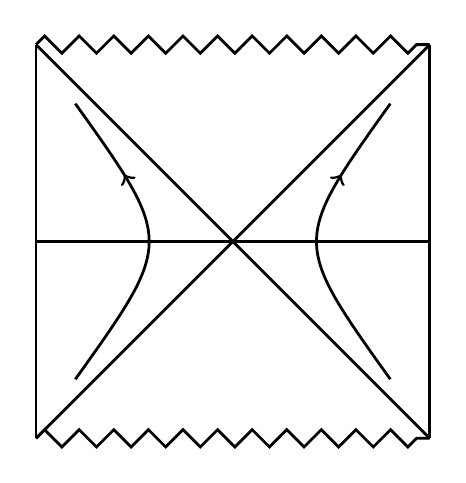}
\caption{}
\label{fig:eternal-two-BH}
\end{subfigure}
\caption{Two different interpretations of the extended AdS-Schwarzschild spacetime. Left: thermofield description of one single black hole; the boost generated by the Hamiltonian propagates upward in the Right Exterior and downward in the Left Exterior. The opposite evolution direction in time implies the fictitious character of the Left Exterior. Right: two black holes in disconnected spaces with a common time.}
\label{fig:eternal-BHs}
\end{figure}

As explained in Ref.~\cite{Maldacena:2013xja}, this state possesses two different interpretations:
\begin{itemize}
\item[(1)] {\it Thermofield interpretation of a single black hole}. In this interpretation, the L sector is fictitious and is introduced as a ``doubling trick'' used in thermofield dynamics. The evolution of the state is given by the Hamiltonian $H_{\rm tf}=H_R-H_L$ where $H_{L,R}$ are the Hamiltonians in the left and right sectors, respectively. The Hamiltonian $H_{\rm tf}$ generates boosts as explained in Fig.~\ref{fig:eternal-one-BH}.

\item[(2)] {\it Two black holes in disconnected spaces}. In this interpretation, the L sector is also physical and hence there are two physical black holes. The Hamiltonian is given by $H=H_R+H_L$. The evolution is explained in Fig.~\ref{fig:eternal-two-BH}.
\end{itemize}

\begin{figure}[H]
\centering
\hspace{-0.5cm}
\begin{subfigure}{0.25\textwidth}
\includegraphics[scale=0.8]{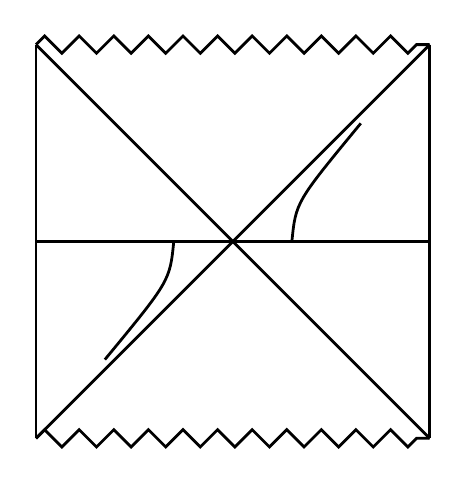}
\caption{}
\label{fig:eternal-one-bubble}
\end{subfigure}
\hspace{+1.8cm}
\begin{subfigure}{0.25\textwidth}
\includegraphics[scale=0.8]{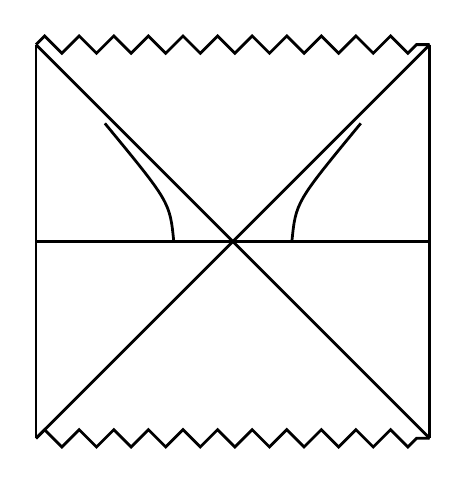}
\caption{}
\label{fig:eternal-two-bubble}
\end{subfigure}
\caption{Left: a static bubble nucleated around a single AdS eternal black hole. The Left Exterior and the left bubble evolves downward and are unphysical. Right: two static bubbles nucleated around two physical AdS black holes. Or one can say that, one bubble is nucleated around the event horizons of a wormhole. }
\label{fig:eternal-bubbles}
\end{figure}

Note that, in whichever interpretation, the individual density matrices on either side are thermal density matrices. Therefore, for a static observer on one side, he/she cannot distinguish from these two interpretations. With these two interpretations and following the analysis given in the last section, it shall be straightforward to see a correspondence between thermal vacuum transitions in the first interpretation and  quantum vacuum transitions in the second interpretation. With the same field system, both kinds of transitions will be described by the same bounce solution in the same Euclidean AdS-Schwarzschild spacetime. The growth of the static bubble after the nucleation is explained in Fig.~\ref{fig:eternal-bubbles}. Since the eternal AdS-Schwarzschild black hole has a holographic description, the bubble nucleation around it could also be described by the CFT on the boundary. This may be left for future work.

\section{A New Paradox from Black Holes?}
\label{sec:paradox}

The discovery of the black hole evaporation has raised many deepest mysteries such as the microscopic origin of the black hole entropy~\cite{Bekenstein:1972tm,Hawking:1974sw}. In particular, by theoretically studying black holes, the information paradox~\cite{Hawking:1976ra} was proposed and remains to be resolved, see Ref.~\cite{Mathur:2009hf} for a pedagogical introduction. The information paradox has been formulated in an extremely sharp version by Almheiri, Marolf, Polchinski and Sully, named as the AMPS firewall paradox~\cite{Almheiri:2012rt}. These paradoxes are intrinsically related to the curious thermal nature of the state observed by the external static observers outside of the hole. 

Here we would like to study a consequence of the correspondence we have constructed so far which may lead to a new paradox related to black holes. The paradox is based on the following assumptions:

\begin{itemize}
\item[(i)] {\it The local field theory is valid near the black hole horizon.} In particular, the state outside of the horizon is thermal viewed from the external static observers as shown by Hawking.

\item[(ii)] {\it The equivalence principle.} In particular, the freely falling observers are in a vacuum state.
\end{itemize}

\begin{figure}
\centering
\includegraphics[scale=0.5]{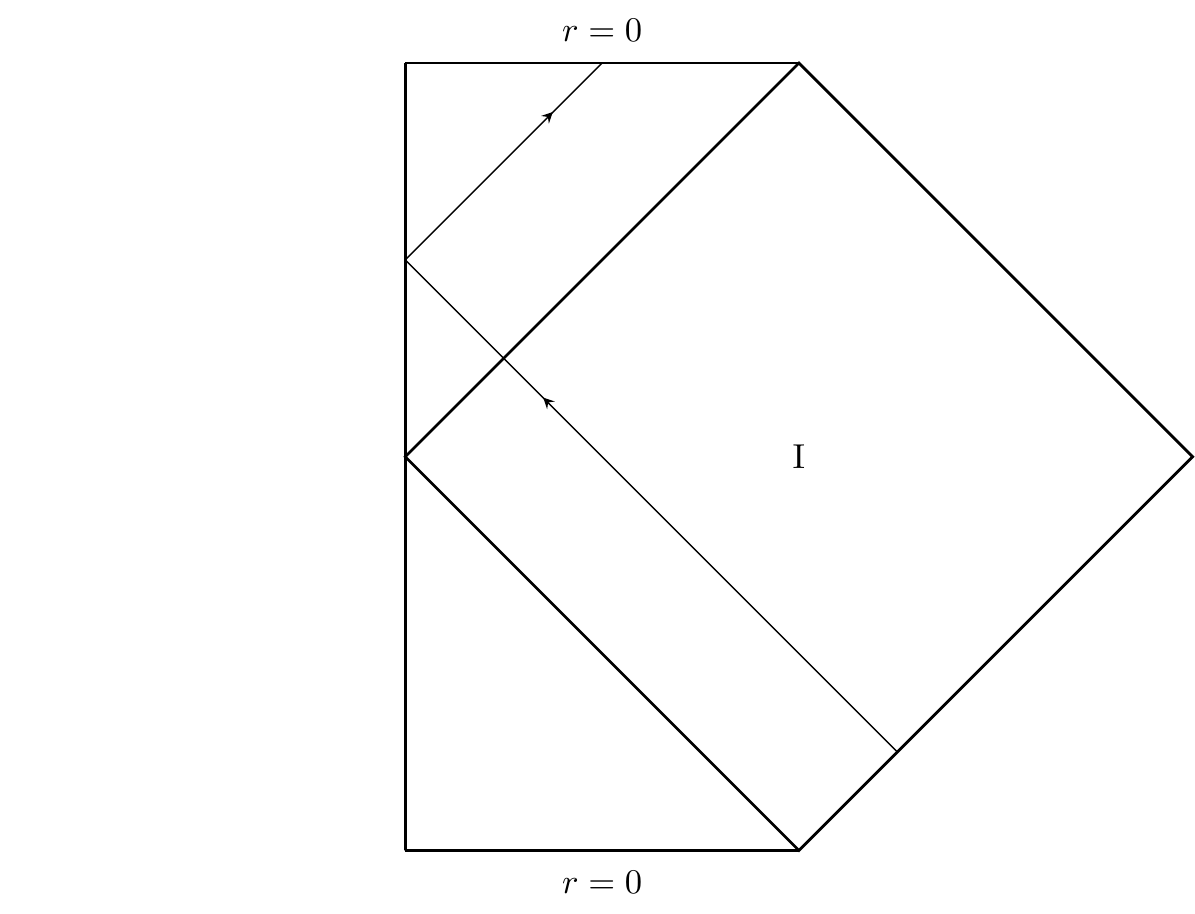}
\caption{Representation of the Schwarzschild spacetime by half of the Kruskal-Szekeres spacetime. The solid line with arrows represents a light beam propagating into the black hole. For realistic black holes, the white hole and its event horizon need to be replaced by the collapsing star.}
\label{fig:S-Penrose}
\end{figure}

The observation is the following. Suppose we have a field system that permits a mild vacuum transition. We can in principle do experiments to measure the probability of bubble nucleation around evaporating Schwarzschild black holes.\footnote{The constructed paradox applies to the eternal AdS black hole as well.} By assumption $(i)$, the static observers are in a thermal ensemble and would measure a result $P_{\rm therm}$ for the thermal bubble nucleations. However, by assumption $(ii)$ the freely falling observers are in a vacuum state and thus see tunneling processes of bubble nucleation. Since the experiments are performed outside of the horizon, the static observers and freely falling observers can compare their results. The freely falling observers must observe the same probability, $P_{\rm tunn}=P_{\rm therm}$. But as we have shown, if the freely falling observers calculate the quantum tunneling probability, they can get a correct prediction only with assuming that they are in a Kruskal-Szekeres spacetime. For the single black hole background, the Kruskal-Szekeres coordinate $X$ needs to be constrained as $X\geq 0$, see Fig.~\ref{fig:S-Penrose}. After the wick rotation $T\rightarrow -i\mathcal{T}$, we only have half a Euclidean Kruskal-Szekeres spacetime. In this case, we will have a different bounce action and tunneling probability. Thus the freely falling observers would get a contradiction between the measured result and the theoretical prediction. 

For Rindler and de Sitter horizons, the ``freely falling observers'' really have another {\it physical} static patch, although it cannot be seen by the ``freely falling observers'' from the patch that our thermal field theory describes. Realizing this, the ``freely falling observers'' in the Rindler spacetime and the de Sitter spacetime can get a correct prediction for the quantum transition. For black holes, it is very hard to imagine that any Schwarzschild black hole is connected with another black hole as described by the Kruskal-Szekeres metric. Note that in this paradox we do not assume an old black hole (after the Page time~\cite{Page:1993wv}) as the information paradox usually does. As long as the event horizon has been formed, we could have such a paradox.

This paradox can be summarized as the following query. The thermal nature that the Rindler and de Sitter horizons present to the static observers on a static patch can be explained by the entanglement between the static patch (say our R-wedge) with another (say our L-wedge). While the black hole horizon has a similar local structure and presents the same thermal nature to the external static observers, what is the origin of such thermal character globally?

This paradox may be related to the ER=EPR conjecture~\cite{Maldacena:2013xja} proposed by Maldacena and Susskind. ER=EPR conjectures that any Einstein-Podolsky-Rosen~\cite{Einstein:1935rr} correlated system is connected by some sort of Einstein-Rosen bridge. Thus somehow, the maximally extended Kruskal-Szekeres spacetime may be indeed realized for any evaporating Schwarzschild black hole, perhaps in some exotic way.

\section{Discussions and Conclusions}
\label{sec:conc}

In this paper, we have clarified the physical meaning of an $O(3)\times O(2)$ bounce solution in the Euclidean Schwarzschild spacetime. We proposed a correspondence between two different interpretations for such bounce. It can be either interpreted as a thermal transition of vacuum in the static region outside of a Schwarzschild black hole or a quantum tunneling in a maximally extended Kruskal-Szekeres spacetime. The thermal and quantum interpretations are given by the static observers or the freely falling observers, respectively. In particular, we found that the Matsubara modes in the thermal interpretation can be mapped to the circular harmonic modes from the $O(2)$ symmetry (from the $O(3)\times O(2)$) in the tunneling interpretation. The transition probabilities in these two interpretations are shown to agree with each other. We also have shown that this correspondence is general and can be applied to the Rindler horizon and the de Sitter horizon.

The correspondence for these two interpretations is due to thermofield dynamics which relates a thermal state to a quantum pure state by doubling the Fock space. Thus the thermal field theory in the static region outside of a Schwarzschild black hole and the (Euclidean version of the) quantum field theory in the maximally extended Kruskal-Szekeres spacetime turn out to live in a same Euclidean spacetime. This correspondence, however, brings a new paradox related to black holes. That is, for freely falling observers, if they calculate the tunneling probability in a Schwarzschild spacetime background, in contrast with the maximally extended Kruskal-Szekeres spacetime, they would not get the agreement with the thermal transition probability obtained by the external static observers.  This paradox is originated from the curious fact that while the black hole horizon shares a similar local structure with other horizons, it is very different in the global structure.

\section*{Acknowledgments}
We would like to thank Bj\"{o}rn Garbrecht and Carlos Tamarit for reading through this manuscript and for their valuable comments. We also thank the anonymous reviewer for suggesting us to extend the discussions to AdS black holes.
This work was supported by the China Scholarship Council (CSC).

\begin{appendix}
\renewcommand{\theequation}{\Alph{section}\arabic{equation}}
\setcounter{equation}{0}

\end{appendix}

\end{document}